\documentclass[11pt,a4paper]{article}
\pdfoutput=1
\usepackage{jcappub}
\usepackage{epsfig} 
\usepackage{epstopdf}
\usepackage{graphicx}
\usepackage{dcolumn}
\usepackage{bm}
\usepackage{color}
\usepackage{calrsfs}
\usepackage{hyperref}
\usepackage{multirow}
\usepackage{float}

\newcommand{\be}{\begin{equation}}
\newcommand{\e}{\end{equation}}
\newcommand{\bear}{\begin{eqnarray}}
\newcommand{\ear}{\end{eqnarray}}

\newcommand{\F}{{\mathcal F}}

\begin{document}

\title{On using large scale correlation of the Ly-$\alpha$ forest and  redshifted 21-cm 
signal to probe HI distribution during the post reionization era}

\author[a]{Tapomoy Guha Sarkar,} 
\author[b]{Kanan K. Datta}  

\affiliation[a]{Birla Institute of Technology and Science, Pilani - 333031, India}
\affiliation[b]{Department of Physics, Presidency University, 86/1 College Street, Kolkata - 700073, India}

\emailAdd{tapomoy@pilani.bits-pilani.ac.in}
\emailAdd{kanan.physics@presiuniv.ac.in}

\abstract{ We investigate the possibility of detecting the 3D cross
  correlation power spectrum of the Ly-$\alpha$ forest and HI 21 cm
  signal from the post reionization epoch. The cross-correlation
  signal is directly dependent on the dark matter power spectrum 
  and is sensitive to the 21-cm brightness temperature and Ly-$\alpha$ forest biases. 
  These bias parameters dictate the strength of anisotropy in
  redshift space. We find that the cross-correlation power spectrum
  can be detected using $400 ~ \, \rm hrs$ observation with SKA-mid (phase 1)
  and a futuristic BOSS like experiment with a quasar (QSO) density of $30 ~
  \rm deg^{-2}$ at a peak SNR of $15$ for a single field experiment at
  redshift $z = 2.5$. We also study the possibility of 
  constraining various bias parameters using the cross power spectrum. We
  find that with the same experiment $1 \sigma$ conditional errors
  on the 21-cm linear redshift space distortion parameter $\beta_T$ and
  $\beta_{\F}$ corresponding to the Ly-$\alpha $ forest are $\sim 2.7
  \%$ and $\sim 1.4 \%$ respectively for $10$ independent pointings of
  the SKA-mid (phase 1). This prediction indicates a significant improvement over
  existing measurements. We claim that the detection of the 3D cross correlation  power spectrum 
  will not only ascertain the
  cosmological origin of the signal in presence of astrophysical foregrounds
   but will also provide stringent constraints on large scale HI
  biases. This provides an independent  probe towards
  understanding cosmological structure formation.  }

\maketitle

\section{Introduction}

Intensity mapping of the neutral hydrogen (HI) distribution using
observations of redshifted 21-cm radiation is a potentially powerful
probe of the large scale structure of the universe and the background expansion history
the post reionization era \cite{poreion1, poreion0, poreion4,
  param2} (also see \cite{hirev3} for a review).  The epoch of
reionization is believed to be completed by redshift $z \sim 6$
\cite{fan2006}. Following this era of phase transition, dense self
shielded Damped Ly-$\alpha$ (DLA) systems contain bulk of the HI gas. 
These DLA systems are believed to be the dominant source of the HI 21-cm
signal in the post reionization era. Mapping the collective HI 21-cm radiation  
without resolving the individual DLAs is expected to yield enormous
astrophysical and cosmological information regarding the large scale
matter distribution, galaxy formation, and expansion history of the 
Universe in the post-reionization era \cite{param2, wyithe08,
  param3, camera13, cosmo14}.

In the same epoch, HI in the dominantly ionized inter galactic medium (IGM)
produces distinct absorption features in the
spectra of background QSOs \cite{rauch98}.  The Ly-$\alpha$ forest,
maps out the HI density fluctuation field along one dimensional
skewers which correspond to QSO sight lines. On suitable large
cosmological scales both the Ly-$\alpha$ forest and the redshifted
21-cm signal are, however, believed to be biased tracers of the
underlying dark matter (DM) distribution \cite{mcd03, bagla2, tgs2011,
  navarro}. Hence, the clustering property of these signals is
directly related to the dark matter power spectrum and the
cosmological parameters.  Like the HI 21-cm signal,  Ly-$\alpha$ forest
observations also find a host of cosmological applications such as
measurement of matter power spectrum \cite{pspec1}, cosmological
parameters \cite{Mandel, pspec2}, limits on neutrino mass
\cite{croft1999}, constraints on the dark energy \cite{cosparam1},
reionization history \cite{reion1} etc. Several Radio interferometric arrays
like the Giant Metrewave Radio Telescope
(GMRT) \footnote{http://gmrt.ncra.tifr.res.in/}, the Ooty Wide Field
Array (OWFA) \cite{saiyad2014}, the Canadian Hydrogen Intensity
Mapping Experiment (CHIME)
\footnote{http://chime.phas.ubc.ca/}, the Meer-Karoo Array Telescope (MeerKAT) 
\footnote{http://www.ska.ac.za/meerkat/}, the Square Kilometer Array
(SKA) \footnote{https://www.skatelescope.org/} are being designed and
are aimed towards observing the background 21-cm radiation for
astrophysical and cosmological investigations. On the other hand,
there has been the recent Baryon Oscillation Spectroscopic Survey
(BOSS) \footnote{https://www.sdss3.org/surveys/boss.php} aimed towards
probing dark energy and cosmic acceleration through measurements
of the large scale structure and the BAO signature in the Ly-$\alpha$
forest \cite{bossdr11}. The availability of high signal to noise ratio
(SNR) Ly-$\alpha$ forest spectra for a large number of QSOs from
the BOSS survey allows 3D statistics to be done with Ly-$\alpha$
forest data \cite{bossdr10, slosar2011}.

Detection of these signals with high statistical significance is
confronted by several observational challenges.  For the HI 21-cm
observations, the signal is extremely weak intrinsically as compared to the
large foregrounds from galactic and extra-galactic sources
\cite{fg1,fg4,fg10}. This inhibits a simple detection. Further,
calibration errors and man made radio frequency interferences plague
the signal. A statistical detection of the signal requires careful
analysis of observational errors and precise subtraction of the
foregrounds \cite{ghosh2011, alonso2014}.  The various difficulties
faced by Ly-$\alpha$ observations include proper modeling and
subtraction of the continuum, flux, incorporating the fluctuations of
the ionizing source, uncertainties in the IGM temperature-density
relation \cite{mac} and contamination of the spectra by metal lines
\cite{bolton}.

The two signals being tracers of the underlying large scale structure
are expected to be correlated on large scales. However foregrounds and
other systematics from two distinct experiments are believed to be
uncorrelated between the two independent observations. Hence, the
cross correlation signal if detected is more likely to ascertain its
cosmological origin. The 2D and 3D cross correlation of the HI 21-cm
signal with other tracers of the large scale structure such as the
Ly-$\alpha$ forest and the Lyman break galaxies have been proposed as a
way to avoid some of the observational issues \cite{tgs5, navarro2}.
The effectiveness of the cross-correlation technique has been
demonstrated by successful detection of the HI 21-cm emission at redshift
$\sim 0.8$ using cross correlations of HI 21-cm maps and galaxies
\cite{masui2013}. It is important to note that the foregrounds in HI 21-cm observations 
appear as noise in the cross correlation and hence, a certain degree
of foreground cleaning is still required for a statistically feasible
detection.

The study of large scale correlation of Ly-$\alpha$ forest
\cite{slosar2011} has reinforced the belief that the Ly-$\alpha$
forest traces the dark matter. Further, the cross correlation of DLAs
and Ly-$\alpha$ forest has been used to measure the DLA bias
\cite{andreu2}. While CMBR observations have been able to precisely
constrain the cosmological parameters, a study of the neutral IGM
requires strong constraints on the bias parameters. These biases are
largely investigated in numerical simulations. Recent measurement of
Ly-$\alpha$ forest parameters using the BOSS survey \cite{slosar2011}
is found to be significantly different from those obtained from
simulation \cite{mcd03}. Precise constraints on these parameters are
extremely important towards understanding the nature of clustering of
the IGM and the physics of structure formation. This motivates us to
investigate the possibility of  measuring large scale HI
bias using the cross-correlation of the Ly-$\alpha$ forest with 21-cm
signal from the post-reionization epoch.

In this paper we first consider the 3D cross power spectrum of the HI
21-cm maps and large scale Ly-$\alpha$ forest. We discuss the
possibility of detecting the signal using the upcoming SKA-mid phase1 (SKA1-mid) like
telescopes and future Ly-$\alpha$ forest surveys with very high QSO
number densities. Finally we consider the possibility
of estimating cosmological parameters using such measurements.  The
fiducial model is chosen to be the $\Lambda \rm{CDM}$ spatially flat
cosmology with parameters taken from the WMAP 7 \cite{wmap7data}. The
cosmological parameter $\Omega_{\Lambda}$ is the only free parameter
for a flat $\Lambda \rm{CDM}$ model. We choose this along with the
redshift space parameters for the 21-cm signal and Ly-$\alpha$
observations. Further, the global amplitude of the cross-power
spectrum is sensitive to the HI neutral fraction in the post
reionization epoch and is chosen as a free parameter. Noting that
cosmological parameters are well constrained from CMBR data, our
primary focus in this work is to investigate how the upcoming
observations may put constraints on the bias parameters for
the Ly-$\alpha$ forest and 21-cm signal.

\section{Formulation} 

\subsection{The Ly-$\alpha$ forest and redshifted 21 cm signal from the post-reionization epoch}

In the post reionization epoch, small fluctuations in the HI density
field in the IGM which is largely ionized, reveal as distinct
absorption features in the spectra of background QSOs known as the
Ly-$\alpha$ forest.  Here, the quantity of observational interest is
the transmitted flux $\F$ through the Ly-$\alpha$ forest. The gas is
believed to trace the underlying dark matter distribution \cite{mcd03,
  slosar2011} on large scales where pressure plays a minor 
role. The neutral fraction is also assumed to be maintained at a
constant value in the IGM owing to photo-ionization equilibrium. This
leads to a power law temperature-density relation \cite{reion,
  gnedin1}. These assumptions are incorporated in the fluctuating
Gunn-Peterson approximation \cite{bidav} relating the
transmitted flux to the dark matter over-density $\delta$ as, 
\be
\frac{\F}{{\bar{\F}}} = {\rm exp}(-\tau) = {\rm exp}\left [- A ( 1 + \delta )
  ^{ 2 - 0.7(\gamma -1)} \right ] 
  \e 
  where $(\gamma -1)$ denotes the
slope of the power law temperature-density relation \cite{gnedin1,
  mac} and is sensitive to the reionization history of the Universe.
The parameter $A \sim 1$ \cite{bolton} varies with redshift and
depends on a host of astrophysical and cosmological parameters, like
the photo-ionization rate, IGM temperature, and parameters controlling
the background cosmological evolution \cite {pspec1}.  It is
however reasonable to assume that $\delta_{\F} = \left ( \F/ \bar{\F}
- 1 \right ) \propto \delta$ with the assumption that the Ly-$\alpha$
forest spectrum has been smoothed over some reasonably large length
scale \cite{vielmat, slosar1}. This linearized relation
facilitates analytic computation of the statistical properties of
$\delta_{\F}$.  We note that dominant corrections to this on small
scales come from peculiar velocities.

The range of redshifts that can be probed using the Ly-$\alpha$ forest
can also be probed using the HI 21-cm emission signal.  However,
unlike the Ly-$\alpha$ forest which arises from the  low density HI
residing in the IGM, the 21-cm emission from the same epoch will be
dominated by DLAs which are believed to contain most of the HI during
the post reionization era. Nevertheless, HI 21-cm signal is likely to
trace the underlying DM distribution on large scales of our interest. 
We use $\delta_{T}$ to denote the redshifted 21-cm brightness
temperature fluctuations.  

We write $\delta_{\F}$ and $\delta_{T}$ in Fourier space as
\be
 \delta_{a}({\bf r}) = \int \ \frac{d^3
   {\bf{k}}}{(2\pi)^3} \ e^{i {\bf{k}}. {\bf r}}  \Delta_{a}({\bf{k}}) \,.
\label{eq:deltau}
\e where $a={\F}$ and $T$ refer to the Ly-$\alpha$ forest
transmitted flux and 21-cm brightness temperature respectively.  With
all the assumptions discussed above and incorporating the effect of
peculiar motion through the redshift space distortion, we may write
\be 
\Delta_{a}({\bf{k}}) = C_{a} [1 + \beta_{a} \mu^2]
\Delta({\bf{k}}) \, 
\e 
where $\Delta({\bf{k}})$ is the dark matter
density contrast in Fourier space and  $\mu$ is the cosine of the angle
between the line of sight direction ${\bf \hat {n}}$ and the wave
vector ($ \mu = {\bf \hat{ k} \cdot \hat{n}}$). $\beta_{a}$ is the linear redshift 
distortion parameter.

For the 21-cm brightness temperature field we have
\be 
C_{T} = 4.0 \, {\rm {mK}} \,
b_{T} \, {\bar{x}_{\rm HI}}(1 + z)^2\left ( \frac{\Omega_{b0}
  h^2}{0.02} \right ) \left ( \frac{0.7}{h} \right) \left (
\frac{H_0}{H(z)} \right) 
\e 
where ${\bar{x}_{\rm HI}}$ is the mean neutral fraction.  The neutral hydrogen
fraction is assumed to be a constant with a value $ {\bar{x}_{\rm
    HI}} = 2.45 \times 10^{-2}$ obtained from the measurement of $\Omega_{gas} \sim 10^{-3}$
\cite{xhibar, xhibar2, zafar2013, noterdaeme}.  
In the case of HI 21-cm signal the parameter $\beta_{T}$, which is known as linear redshift
distortion parameter, can be written as the ratio between the growth
rate of linear perturbations $f(z)$ and the HI bias $b_T$. The bias function $b_T(k, z)$ has 
an intrinsic scale dependence below the Jeans scale and an implicit scale dependence 
arising from the fluctuations in the ionizing background
\cite{poreion0}. Moreover, it has been shown \cite{marin} that this
bias is a monotonically growing function of redshift. The assumption
of linear bias is supported by numerical simulations \cite{bagla2,
  tgs2011} which indicate that over a wide range of scales,  a constant
bias model is adequate to describe the distribution of neutral gas for
$z < 3$. Recent measurement of DLA bias is also consistent with the
constant bias model \cite{andreu2}. We find that $f(z) \approx 1$ at the fiducial redshift of interest $ z = 2.5$.  
We  adopt a constant bias $b_{T} = 2$ which is consistent with recent results from
numerical simulations of HI 21-cm signal in the post-reionization epoch
\cite{bagla2, tgs2011, navarro}.  This gives the value of the parameter $\beta_{T} \approx 0.5$.

The linear distortion parameter for the Ly-$\alpha$ forest, denoted by
$\beta_{\F}$, can not be interpreted in the same way as
$\beta_{T}$. This is because of the non-linear relationship between the observed Ly-$\alpha$ transmitted flux
and the underlying DM density field
\cite{slosar2011}. The bias factor for the forest is the bias of
  the contrast of the fluctuations in the flux and is not same as the
  HI bias. Unlike the HI 21-cm signal, the parameters
  $(C_{\F}, \beta_{\F})$ are independent of each other and are
  dependent on the model parameters $A, \gamma$ and the flux probability distribution function (PDF) of
  the Ly-$\alpha$ forest. In the absence  of
  primordial non-gaussianity, fluctuations in the Ly-$\alpha$ flux can be well described 
  by a linear theory with a scale independent bias on large scales. This is supported by 
   numerical simulations \cite{mcd03}. The values of $\beta_{\F}$
  obtained in particle-mesh (PM) simulations demonstrate that simulations with lower
  resolution (larger smoothing length) yield lower
  values of $\beta_{\F}$. The smoothing scale is ideally set by the
  Jean's scale which is sensitive to the temperature history of the
  IGM. 

 We adopt an approximate values $(C_{\F}, \beta_{\F}) \approx ( -0.15,
 1.11 )$ from the numerical simulations of Ly-$\alpha$
 forest \cite{mcd03}. We note that for cross-correlation studies the
 Ly-$\alpha$ forest has to be smoothed to the resolution of the HI 21 cm
 frequency channels. On these smoothed scales the linear bias model
 is well tested in simulations. Results of full hydrodynamical
 simulations are expected to yield information at smaller scales which
 are not necessary for the present analysis.  We, however, note that
 these bias values have large uncertainties owing to the lack of
 accurate modeling of the IGM.  The redshift space distortion parameter
 $\beta_{\F}$, is also sensitive to the probing redshift.  The large
 scale correlation of Ly-$\alpha$ transmitted flux from BOSS survey
 \cite{slosar2011} shows that the parameters $(C_{\F}, \beta_{\F})$ are 
 significantly different from the above values obtained from
 simulations. However it has been suggested that metal line contamination and
 correct treatment of DLAs may explain this discrepancy. In this paper
 we shall stick to the values obtained from simulation \cite{mcd03}.

\subsection{Cross-correlation power spectrum}

The possibility that both the Ly-$\alpha$ forest and the HI 21-cm signal from the
post reionization epoch trace the underlying dark matter density
field on large scales, motivates us to investigate their cross-correlation signal
\cite{tgs5}. Though the respective auto-correlation power spectra can
independently put constraints on various astrophysical and cosmological parameters
there are several advantages of cross-correlating the signals.

 The main advantage of cross-correlation is that the issue of
 foregrounds and other systematics can be coped with greater ease as
 compared to the auto correlation.  Even the smallest foreground
 residual will plague the auto-correlation signal. The cosmological origin of the
 signal can only be ascertained if it is detected with
 statistical significance in cross-correlation.   Further, a joint analysis of two data sets
 would involve not only the individual auto-correlation but also the
 cross-correlation information.  Sometimes the two independent probes
 focus on specific complimentary Fourier modes with high SNR whereby
 the cross signal takes advantage of both the probes simultaneously. This has been 
studied in the context of BAO where, owing to the difference in values of the parameters $\beta_{T}$ and $ \beta_{\F}$ 
the two probes have different sensitivities to radial and transverse clustering \cite{tgs6}. 

It is true that if the observations of the independent probes are perfect measurements
no new information can be obtained from the cross correlation.
However,  the first generation measurements of the HI 21 cm signal 
are expected to be noisy and shall have systematic errors. For a detection of the 21 cm signal 
these measurements can in principle
 be cross-correlated against a high SNR Ly-$\alpha$ forest signal for 
cosmological investigations which may not be possible with the low quality auto correlation analysis.

We consider the power spectrum of 21-cm signal, the Ly-$\alpha$ forest and the cross 
correlation power spectrum in three dimensions. The general 3-D power spectra for the two
fields are defined as \be \langle \ \Delta_{{a}}({\bf{k}})
\Delta^{*}_{b}({\bf{k}'}) \ \rangle = (2 \pi)^3 \delta^3 ({\bf{k}}
- {\bf{k}'}) P_{ab} ({\bf{k}})
\label{eq:fluxps3}
\e 
where $a , b $ can be $ \F$ and $T$. In general the power spectrum can be written in 
redshift space as,
\be
P_{ab} ({\bf{k}})=C_aC_b(1+\beta_a \mu^2)(1+\beta_b \mu^2) P({\bf{k}})
\e
where  $P({\bf{k}})$ is the matter power spectrum. The auto-correlation power spectrum
corresponds to $a = b$ and the cross-correlation power spectrum corresponds to $a \neq b$.

 The cross-correlation can be computed only in the region of overlap
 between the observed Ly-$\alpha$ forest and 21-cm fields.   However, we note that 
 in real observations the Ly-$\alpha$ forest surveys are likely
 to cover much larger volume than the single field radio
 observations of the HI 21-cm signal.  We consider such an overlap volume
 $\mathcal{V}$ consisting of a patch of angular extent $ \theta_a
 \times \theta_a$ on the sky plane and of thickness $L$ along the line
 of sight direction. We consider the flat sky approximation. This
 amounts to writing the comoving separation vector ${\bf r}$ as \be
 {\bf r} = r_{\nu} {\vec \theta} + {\bf \hat n} \frac{dr}{d \nu} \nu
 \e where $ r_{\nu}$ is the comoving distance corresponding to the
 observing frequency $\nu$ and ${\vec \theta}$ is a 2D vector on the
 sky plane. If $B$ denotes the bandwidth of the 21-cm observation, we
 have $ L = B dr/d\nu$ and $\mathcal{V} = r_{\nu}^2 \theta_a^2 L$. The
 observed 21-cm signal (after significant foreground cleaning) in
 Fourier space is written as
\be 
{\Delta}_{{T} o}({\bf{k}}) = \Delta_{T}({\bf{k}}) +
\Delta_{NT}({\bf{k}}), 
\e 
where $\Delta_{NT}$ is the corresponding
noise. The radio observations measure visibilities as a function of
the 2D baseline vector ${\bf U}$ and frequency $\nu$. We have ${\bf k}
= ( {\bf k}_{\perp}, k_{\parallel}) = ( 2 \pi {\bf U}/ r_{\nu} , 2 \pi
\tau d \nu /dr) $ where $\tau$ is the Fourier conjugate variable
corresponding to $\nu$. The 21-cm brightness temperature fluctuations
in Fourier space are closely related to the measured Visibilities.

The Ly-$\alpha$ flux fluctuations are written as a field in the 3-D
space as $\delta_{\F}({\bf{r}})$. In reality one has the observed
quantity $\delta_{{\F}o}({\bf{r}})$ which consists of the continuous
field sampled along skewers corresponding to QSO sight lines. We,
hence have $\delta_{{\F} o}({\bf{r}})=\delta_{\F}({\bf{r}}) \times
\rho({\bf{r}}) $, where the sampling function $\rho({\bf{r}})$ is
defined as 
\be 
\rho({\bf{r}}) =\frac {\sum_i w_i
  \ \delta_D^{2}({\bf{r}}_{\perp} - {\bf{r}}_{\perp a}) }{L \sum_i
  w_i} 
  \e 
  and is normalized to unity ( $\int dV \rho({\bf{r}}) = 1$ ).
The index $'i'$ goes up to $N_{Q}$, the total number of QSOs
considered. The weights $w_i$ introduced in the definition of $\rho$
are chosen so as to minimize the variance. The suitable choice of
the $w_i$ takes care of the fact that the pixel noise for each of the
QSO spectra are in principle different. We have, in Fourier space, 
\be {\Delta}_{{\F} o}({\bf{k}}) =
\tilde{\rho}({\bf{k}}) \otimes {\Delta_{\F}}({\bf{k}}) +
\Delta_{N\F}({\bf{k}}),
\e 
where $\tilde{\rho}$ is the Fourier transform
of $\rho$,  $ \otimes $ denotes a convolution and $
\Delta_{N\F}({\bf{k}})$ denotes a noise term.

 We define the cross-correlation estimator $\hat{\mathcal{E}}$ as 
 \be
 \hat{\mathcal{E}} = \frac{1}{2} \left[ {\Delta}_{{\F} o}({\bf{k}})
   {\Delta}^{*}_{{T} o}({\bf{k}}) + {\Delta}^{*}_{{\F} o}({\bf{k}})
   {\Delta}_{{T} o}({\bf{k}}) \right]. 
   \e 
   We are interested in the
 statistical properties of this estimator. Using the definitions of
 ${\Delta}_{{\F} o}(\bf{k})$ and ${\Delta}_{T o}({\bf{k}})$, we obtain
 the expectation value of $\hat{\mathcal{E}}$. Simple algebraic
 manipulation yields 
 \be \langle \ \hat{\mathcal{E}} \ \rangle = P_{\F
   T}({\bf{k}}). 
   \e 
   Thus, the estimator is unbiased and its expectation
 value faithfully returns the quantity we are probing, namely the 3-D
 cross-correlation power spectrum $ P_{\F T} ({\bf{k}})$.  We have
 assumed that the different noises are uncorrelated.  Further, we note
 that the QSOs are distributed at a redshift different from rest of
 the quantities and hence $\tilde{\rho}$ shall be uncorrelated with
 both $\Delta_T$ and $\Delta_{\F}$.  
 
 \subsection{Variance of the estimator and Fisher Matrix analysis}
 
 The variance of the estimator
 $\hat{\mathcal{E}}$, defined as, $ \sigma_{\hat{\mathcal{E}}}^2 =
 \langle {\hat{\mathcal{E}}}^2 \rangle - {\langle \hat{\mathcal{E}}
   \rangle }^2$ gives
\begin{eqnarray}
\sigma_{\hat{\mathcal{E}}}^2 = \frac{1}{2} P_{\F T}({\bf k})^2 +
&&\frac{1}{2} \left [ P_{\F \F}({\bf k}) + P^{\rm {1D}}_{\F
    \F}(k_{\parallel}) P^{\rm {2D}}_{w} + N^{}_{\F} \right ]
  \nonumber \\ \times \, && \left[ P_{T T }({\bf k}) + N_{T} \right].
\label{eq:sigma}
\end{eqnarray}
The quantity $ P^{\rm {1D}}_{\F \F}(k_{\parallel})$ is known as the
  aliasing term and is the usual 1-D Ly-$\alpha$ flux power spectrum
  of the individual spectra given by
  \be P^{\rm {1D}}_{\F
    \F}(k_{\parallel}) = \frac{1}{(2\pi)^2} \int d^2{\bf k}_{\perp}
  P_{\F \F}({\bf k}). 
  \e 
  The quantity $ P^{\rm {2D}}_{w} $ denotes the power
  spectrum of the weight function. The quantities $ N_{T}$ and $
  N^{}_{\F}$ denote the effective noise power spectra for the 21-cm
  and Ly-$\alpha$ observations respectively. Writing ${\bf k} = ( {\bf k}_{\perp}, k_{\parallel}) $ with $| {\bf
  k}_{\perp}| = k_{\perp} = k \sin \theta $ and $k_{\parallel} = k
\cos \theta $, we have \be \delta P_{\F T}({\bf k}) = \frac
     {\sigma_{\hat{\mathcal{E}}}}{\sqrt{N_m(k, \theta)}} \e where
     $N_m(k,\theta)$ is number of observable modes in between $k$ to
     $k+dk$ and $\theta$ to $\theta+d \theta$ given by 
     \be N_m (k,
     \theta) =\frac{2 \pi k^2 \mathcal{V} \sin \theta \, dk \, d
       \theta}{(2 \pi)^3}.  
       \e 
       For the Ly-$\alpha$ forest one may choose the weights $w_i$ of the
inverse variance form \cite{kaipea}. However, an uniform weighing scheme
suffices when most of the spectra are measured with a sufficiently
high SNR \cite{mcquinnwhite}. This gives $ P^{\rm {2D}}_{w} = \frac{1}{\bar{n}}$,
where $\bar{n}$ is the 2D density of QSOs ($\bar{n} = N_{Q}/
{\mathcal{A}}$).  We assume that the variance $\sigma^2_{{\F} N}$ of
the pixel noise contribution to $\delta_{\F}$ is the same across all
the QSO spectra whereby we have $N_{\F} = {\sigma}^2_{{\F}
  N}/\bar{n}$ for its noise power spectrum.  In arriving at equation
(\ref{eq:sigma}) we have ignored the effect of QSO clustering. In
reality, the distribution of QSOs is expected to exhibit clustering.  
The clustering would enhance the term $ \left (P^{\rm
  {1D}}_{\F \F}(k_{\parallel}) P^{\rm {2D}}_{w} + N^{}_{\F} \right )$
  in equation (\ref{eq:sigma}) by a factor $ \left ( 1 + {\bar n}
  C_{Q}({\bf k }_{\perp}) \right) $, where $C_{Q}({\bf k }_{\perp})$
  is the angular power spectrum of the QSOs.  However, for the
  QSO surveys under consideration, the Poisson noise dominates over
  the clustering term and the latter may be ignored.

We consider a radio-interferometric measurement of the 21-cm signal
whereby the instrumental sensitivity per ($k$) mode to the redshifted
21-cm power spectrum at an observed frequency $\nu = 1420/(1 + z) {\rm
  MHz}$ can be calculated using the relation \cite{fg2} 
\be
N_{T} (k,\nu)=\frac{T_{sys}^2}{B t_0} \left ( \frac{\lambda^2}{A_e}
\right)^2 \frac{r_{\nu}^2 L}{n_b(U,\nu)}. 
\label{eq:noise-error}
\e 
Here, $T_{sys}$ denotes the system temperature. $B$ and $L$ are, as
described before, the observation bandwidth and comoving length
corresponding to the bandwidth $B$ respectively, $t_0$ is the total
observation time, $r_{\nu}$ is the comoving distance to the redshift
$z$ , $n_b(U,\nu)$ is the number density of baseline $ U$, where
$U=k_{\perp}r_{\nu}/2 \pi$, and $A_e$ is the effective collecting area
for each individual antenna. We may write $n_b(U,\nu)$ as 
\be
n_b(U,\nu)=\frac{N(N-1)}{2}f_{2D}(U,\nu), 
\e 
where $N$ is the total
number of antennae in the radio array and $f_{2D}(U,\nu)$ is the
normalized baseline distribution function which follows the normalization
condition $\int d^2 {\bf U} f_{2D}(U,\nu)=1$.  The Ly-$\alpha$ forest
flux and the 21-cm signal are modeled using parameters $ \left(
C_{\alpha} , \beta_{\alpha}\right ) $.  The quantity $A = C_{T}
C_{\F}$ appears as a single overall constant and is largely uncertain.
We note that the parameter $C_{\alpha}$ also includes the bias parameter
$b_{\alpha}$. Additionally, we consider the possibility of constraining
this amplitude, two linear redshift space distortion parameters i.e, $\beta_{T}$ and
$\beta_F$ along with the cosmological parameter $\Omega_{\Lambda}$
assuming a flat $\Lambda$CDM model i.e, $\Omega_{\Lambda}+\Omega_{m}=1$. We
label these $4$ parameters as $\lambda_{r} $.  The Fisher matrix used
for estimation of parameters is given by the $4 \times 4$ matrix 
\be
F_{r s } = \int \int \frac{1}{ \sigma_{\hat{\mathcal{E}}}^2}
\left(\frac{\partial P_{\F T}}{\partial \lambda_r}\right)
\left(\frac{\partial P_{\F T}}{\partial \lambda_s}\right) \frac {2 \pi
  \mathcal{V} k^2 dk d\mu }{(2 \pi)^3}, 
  \e 
  where $\sigma_{\hat{\mathcal E}}$
is obtained from equation (\ref{eq:sigma}).  The marginalized error on the $i ^{th}
$ parameter $\Delta \lambda_i$ is given by the Cramer-Rao bound;
$\Delta \lambda_i = \sqrt {F^{-1}_{i i}}$.

This gives the theoretical bound for the
error in a given parameter. If the errors are correlated, then in the
space of the parameters we shall have error contours corresponding to
the significance at which statistical detection is sought. Assuming the
Cramer-Rao bound, the error contours are expected to be elliptic whose 
areas measure the figure of merit, and the orientation of the principal axis
measures the strength of correlation between the parameters.

\section{Results} 

\begin{figure}
\begin{center}
\includegraphics[height=7cm, width=7cm, angle=-90]{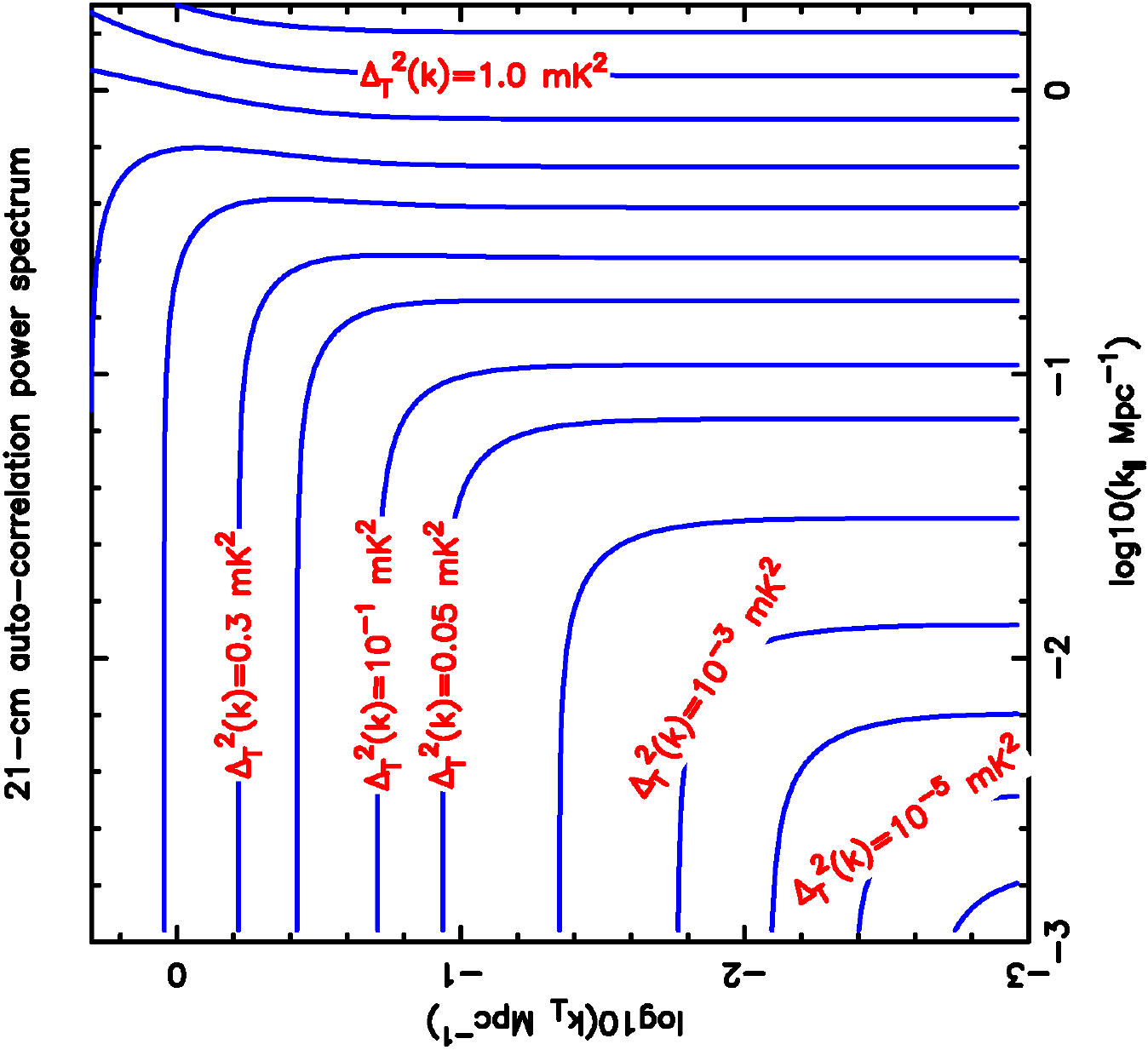}
\includegraphics[height=7cm, width=7cm, angle=-90]{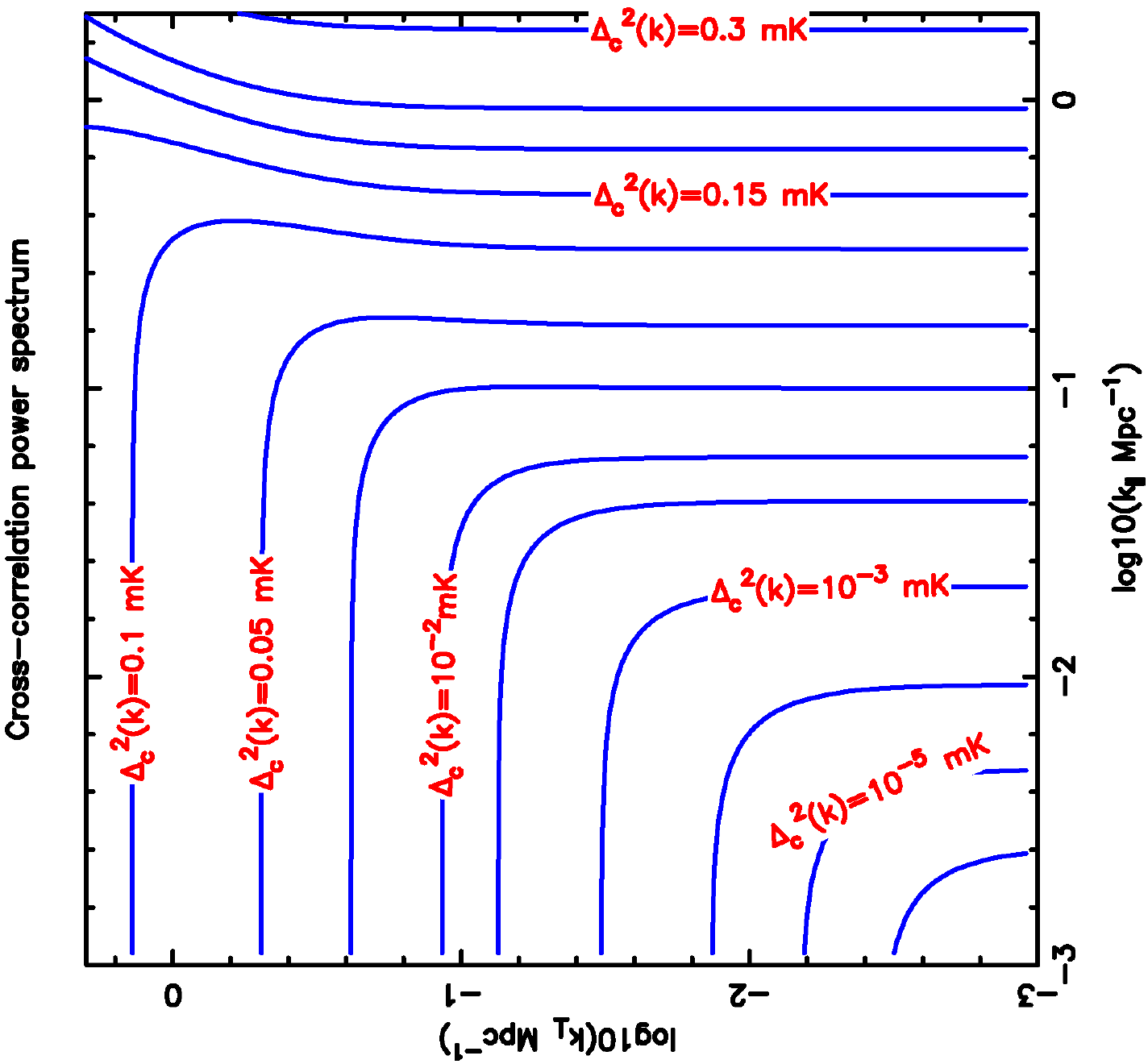}
\end{center}
\caption{Figure shows the 3D power spectrum in redshift space at a fiducial 
redshift $z = 2.5$. The left panel shows the 21-cm  power spectrum 
$\Delta_T^2 = k^3 P_{TT}({\bf k})/ 2 \pi^2$ and the right panel shows the 
cross-correlation power spectrum $ \Delta_C^2 =  k^3 P_{T \F}({\bf k})/ 2 \pi^2$. The 
redshift space distortion  appears as deviations from spherical symmetry of the power 
spectrum.}
\label{fig:sigHips}
\end{figure}


\subsection{Power spectrum and its detection}
The left panel of the figure (\ref{fig:sigHips}) shows the dimensionless 21-cm power spectrum
($\Delta^2_T(k_{\perp}, k_{||})=k^3 P_T(k_{\perp}, k_{||})/2 \pi^2$)
in redshift space at the fiducial redshift $z = 2.5$. In the range of
modes of interest $ 10^{-2} \lesssim k \lesssim 1\, {\rm Mpc}^{-1}$ the
  signal varies in the range $ 10^{-3} \lesssim \Delta_T^2 \lesssim 1\,
  {\rm mK^2}$. We find that in the plane of $k_{||}$ and $k_{\perp}$ the power spectrum 
  is not circularly symmetric. The
  degree of asymmetry is sensitive to the redshift space distortion
  parameter. Thus, for a given k-mode the power spectrum differs for
  different sets of $(k_{\perp}, k_{||})$ values.  The right panel of figure (\ref{fig:sigHips})
  shows the 21-cm and Ly-$\alpha$ cross-power
  spectrum. The departure from spherical symmetry is noted here as
  well. However, the magnitude and scale dependence of the distortion
  is different from the 21-cm auto power spectrum owing to the
  difference between the linear distortion parameters $\beta$ for
  Ly-$\alpha$ forest and the 21-cm signal. We note that anisotropies in the
  power spectrum are determined by linear distortion parameters
  $\beta$.  The quantity $\beta_F$ is $\sim 3$ times larger than
  $\beta_T$. Hence, the cross-correlation power spectrum is more anisotropic than the 
  HI 21-cm auto correlation power spectrum. The cross correlation signal varies
  in the range $ 10^{-4} \lesssim \Delta_c^2 \lesssim 10^{-1}\, {\rm mK}$ in the
  $k$-range $10^{-2} \lesssim k \lesssim 1\, {\rm Mpc}^{-1}$.

We have chosen a fiducial redshift of $z=2.5$ for our analysis. This
is justified since the QSO distribution is known to peak in the
redshift range $ 2 < z < 3$.  Further, to avoid metal line
contamination and the effect of the QSO stromgen sphere, only a
part of the QSO spectra is to be considered. At the fiducial
redshift this corresponds to approximately a redshift band $\Delta z
\sim 0.4$. The cross-correlation can however only be computed in the
region of overlap between the 21-cm signal and the Ly-$\alpha$ forest
field. This is dictated by whichever is smaller - the band width of
the 21-cm signal or the redshift range over which one has the
Ly-$\alpha$ spectra.

The details of the radio interferometer specifications used here,
other than the baseline distribution function can be found in a recent
paper \cite{navarro}. In brief, we consider a radio interferometric
array for the 21-cm observations mimicking the SKA1-mid. The SKA1-mid
is one of the three different instruments that will be built as a part
of the SKA telescope. Although these specifications may undergo
changes,  we use the specifications provided in the `Baseline
Design Document'\footnote{available here:
  https://www.skatelescope.org/key-documents/}.

  We now describe the telescope specifications used for our analysis. We consider an  operational frequency
range of $350$ MHz to $14$ GHz. We assume a  total of $250$ dish
like antennae each of $\sim 15$m diameter \footnote{ A recent document  ``SKA Level 1 Requirements (revision 6)" 
(https://www.skatelescope.org/key-documents/) has indicated that only
  $\sim 50 \%$ of these antennae may be deployed. To achieve the same label of accuracy presented 
  in this work with this degraded design one has to approximately consider $4$ times the total 
  observation time projected in this paper. However, this is a naive scaling 
  and redesigned baseline distribution may change the entire analysis. For example, if the 
  reduction is only of the large baseline antennae then the sensitivities presented here may not be 
  that severely affected. }.  To calculate the
normalized baseline distribution function $f_{2D}(U, \nu)$ we use the
baseline density provided in \cite{navarro2} (blue line in their
Fig. 6). The blue line is essentially proportional to $f_{2D}(U) $ for
a given frequency $\nu$. We use the normalization condition $\int d^2
{\bf U}f_{2D}(U,\nu)=1$ to calculate the normalization factor. We note
that the baseline distribution is centrally condensed with $40\%$,
$55\%$, $70\%$, and $100\%$ of the total antennae are within $0.35$
km, $1$ km, $2.5$ km, and $100$ km radius respectively. We also assume
that there is no baseline coverage below $30$m. Centrally condensed
baseline coverage helps to achieve sufficient power spectrum
sensitivity at large scales. The coverage is poor at very small scales
(large $U$). However, owing to non-linear effects the modeling of the
power spectrum at these small scales is anyway quite incomplete. We
plug everything in Eq. \ref{eq:noise-error} to obtain the required
noise error in the 21 cm power spectrum for a given $(k_{\perp},
k_{\parallel})$. Then, for a given bin $(k_{\perp}+dk_{\perp},
k_{\parallel}+d k_{\parallel})$ we calculate the total number of
independent modes $N_c$ and reduce the noise rms. by a factor of
$\sqrt{N_c}$. We assume $T_{sys}$ to be $30$K for the redshifts $z =
2.5$ corresponding to observational frequency of $405.7$ MHz. We also
consider observations over $32$ MHz bandwidth and a typical antenna
efficiency equal to $0.7$ .

  We note that for a QSO at a given redshift, the region $10,000 \,
  {\rm km \, s^{-1}}$ blue-wards of the QSO's Ly-$\alpha$ emission
  has to be excluded from the Ly-$\alpha$ forest to avoid the QSO's
  proximity effect.  Further, at least 1,000 $\rm km \, s^{-1}$
  red-ward of the QSO's Ly-$\beta$ and O-$\rm VI$ lines may be discarded
  to avoid any confusion with the Ly-$\beta$ forest or the intrinsic
  O-$\rm VI$ absorption.  For example a QSO at a fiducial redshift
  $2.5$, this would allow the Ly-$\alpha$ forest to be measured in the
  redshift range $ 1.96 \leq z \leq 2.39$ spanning an interval $\Delta
  z =0.43$.  It is necessary to consider a survey with a higher QSO
  density for the cross correlation SNR to be competitive with that of
  the 21-cm auto-correlation. We consider a BOSS-like QSO survey with
  a QSO density of $30 \ \rm deg^{-2}$ which are measured at average
  $2\sigma$ sensitivity. Though the QSO surveys cover a large portion
  of the sky, $\sim 10,000 \rm deg^2$, the cross correlation can only
  be computed in the region of overlap of the 21-cm and Ly-$\alpha$ forest
  survey.

\begin{figure}[h]
\begin{center}
\includegraphics[height= 7cm, width= 7cm, angle=-90]{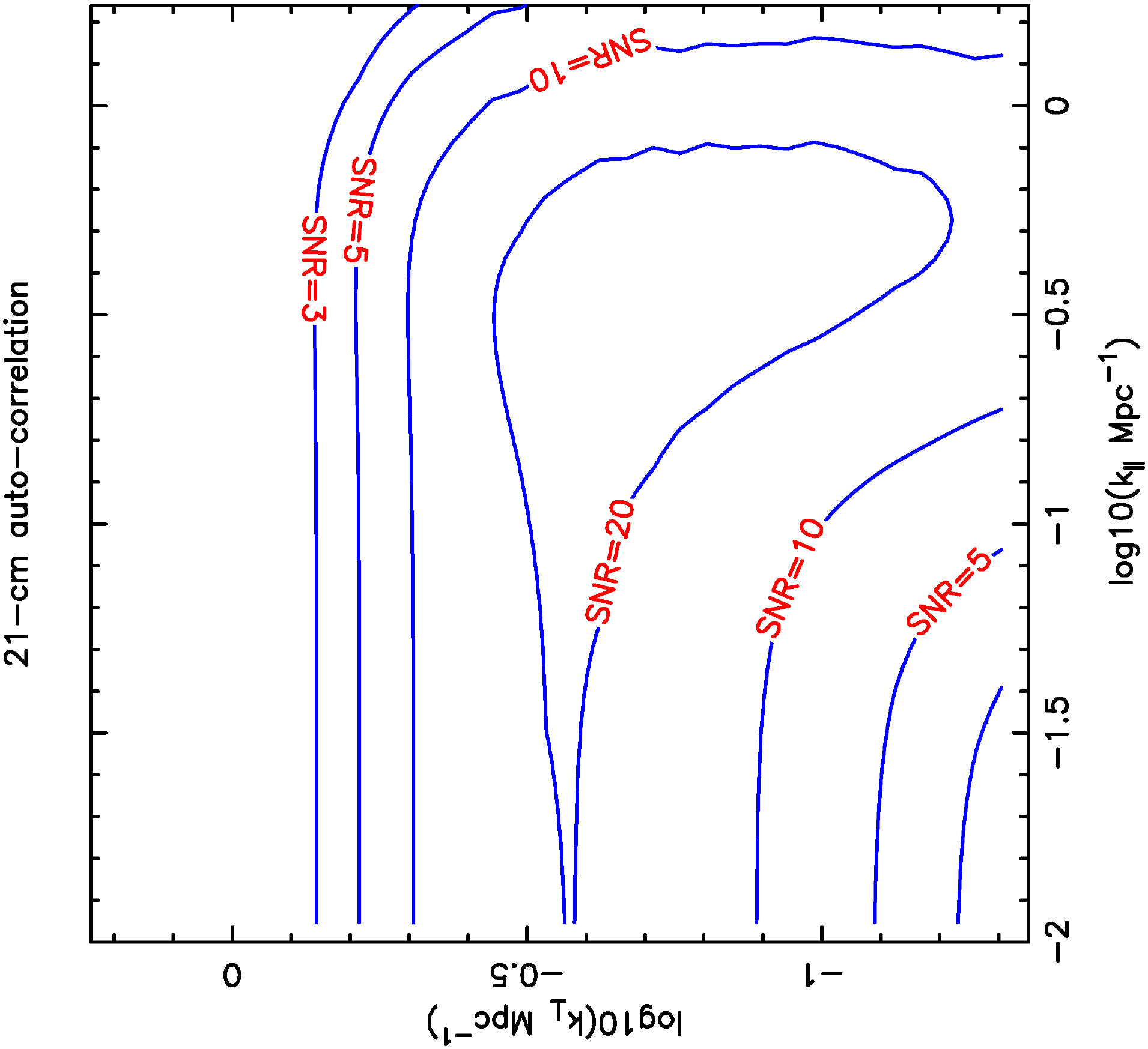}
\includegraphics[height= 7cm, width= 7cm, angle=-90]{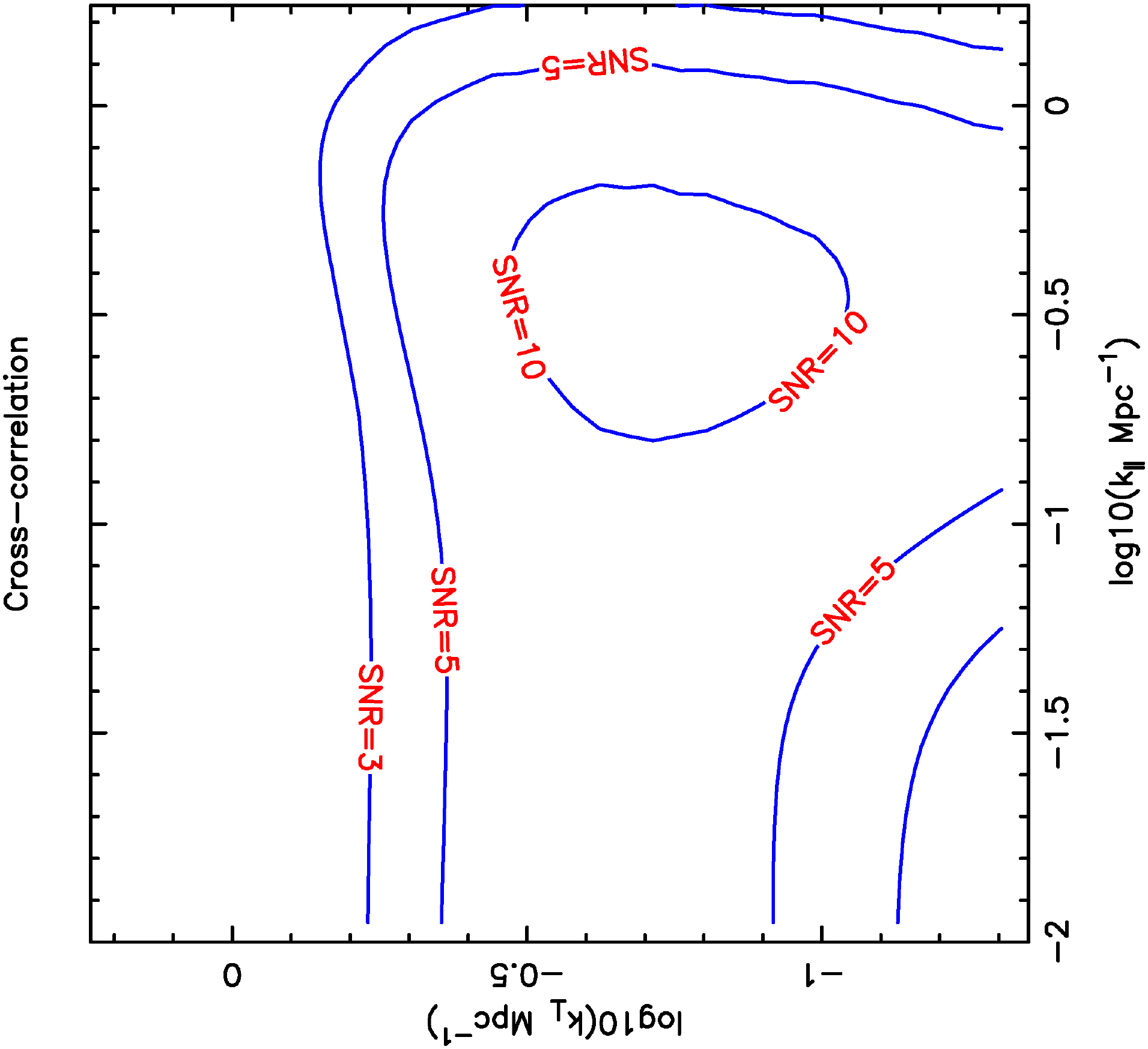}
\end{center}
\caption{SNR contours for the 21-cm auto-correlation 
  (left panel) and the cross correlation (right panel) power spectrum in redshift space 
  at the fiducial redshift $z= 2.5$ for a $400 \rm
  hrs$ observation at $405 \rm MHz$ assuming that complete foreground
  removal is done.}
\label{fig:Hipps-nofg}
\end{figure}

We first consider the idealized situation where the foregrounds in the
21-cm observations are completely absent. This means that a perfect foreground 
subtraction has been achieved. The left panel of the figure
(\ref{fig:Hipps-nofg}) shows the SNR contours for the 21-cm auto
correlation power spectrum for a $400 \rm \, hrs$ observation and total
$32$MHz bandwidth at a frequency $405.7 \rm \, MHz$ corresponding to $z =
2.5$ in this idealized situation. We have taken a bin size to be
$(\Delta k, \Delta \theta)=(k/5, \pi/10)$. The SNR reaches at the peak
($>20$) at intermediate value of $(k_{\perp}, k_{\parallel})=(0.4,0.4)
\, {\rm Mpc^{-1}}$ and falls off at both lower and higher
$k$-values. We find that $5\sigma$ is possible in the range $0.08
\lesssim k_{\perp} \lesssim 0.6 \, {\rm Mpc^{-1}}$ and $0.1 \lesssim
k_{\parallel} \lesssim 1.5 \, {\rm Mpc^{-1}}$. The similar range for
the $10\sigma$ detection is $0.12 \lesssim k_{\perp} \lesssim 0.5 \,
{\rm Mpc^{-1}}$ and $0.2 \lesssim k_{\parallel} \lesssim 1.2 \, {\rm
  Mpc^{-1}}$. At lower values of $k$, the noise is dominated by cosmic
variance whereas, the noise is predominantly instrumental at large
$k$. However, lower cut-off for the $k_{\parallel}$ arises from the
limited bandwidth which we assume to be $32$ MHz in this work. The
presence of redshift space distortion shows up as an asymmetry in the
sensitivity contours in the $(k_{||}, k_{\perp}) $ plane. The enhanced
radial clustering due to redshift space distortion effect manifests as
higher SNR ( say $ > 10$) for a larger range of $k_{||}$ than
$k_{\perp}$.


\begin{figure}[]
\begin{center}
\includegraphics[height= 7cm, width=7cm, angle=-90]{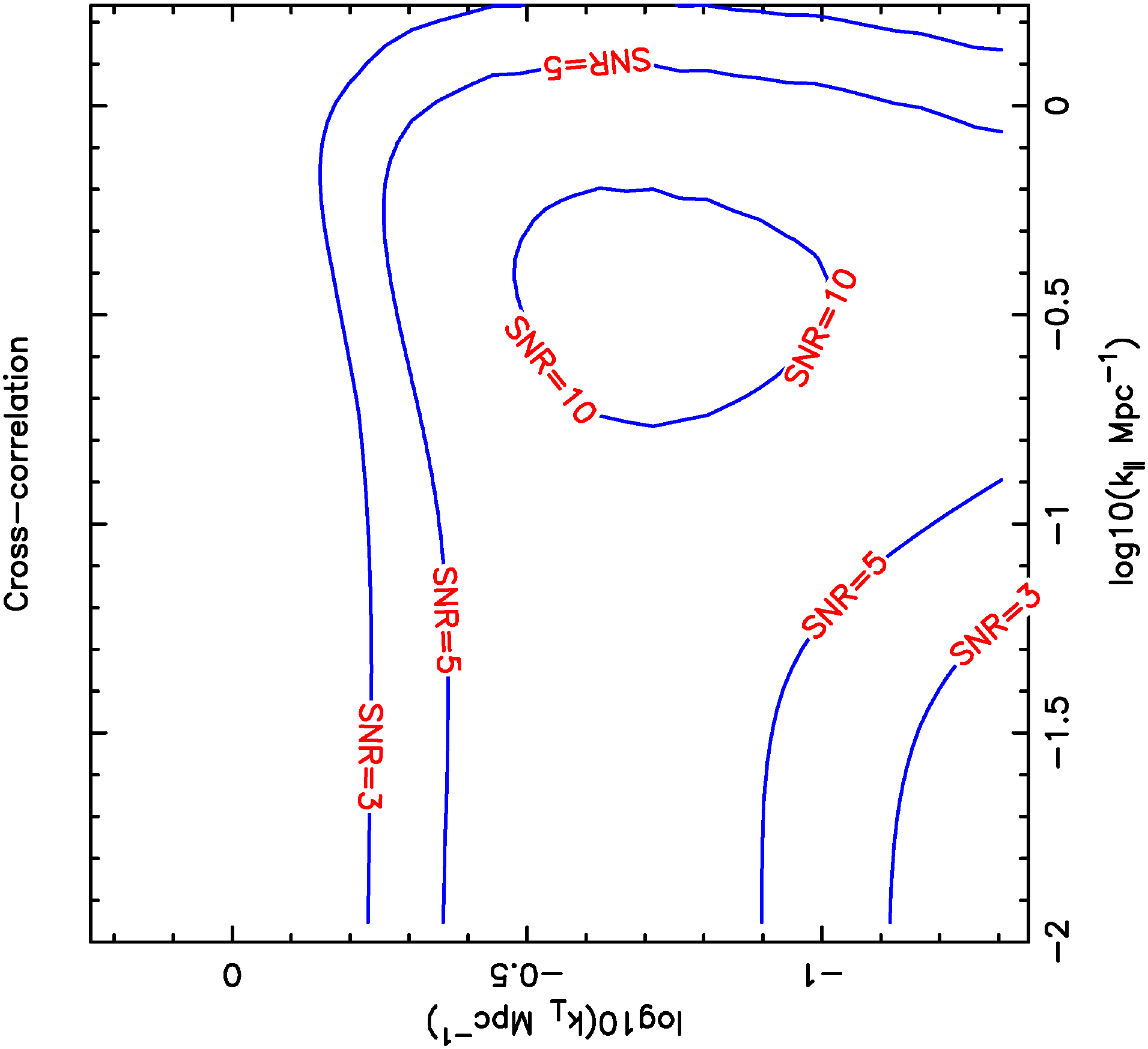}
\includegraphics[height=7cm, width= 7cm, angle=-90]{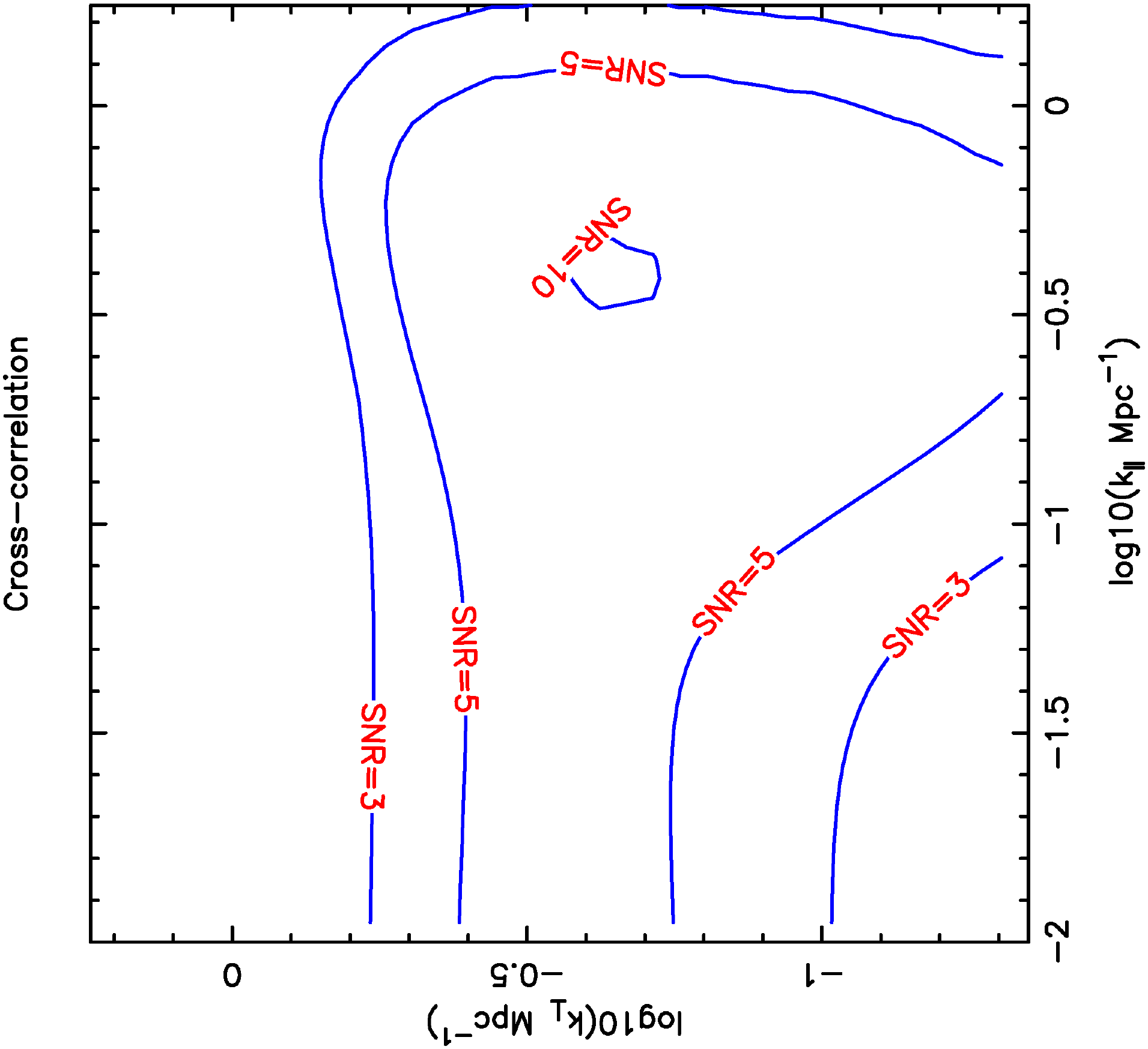}
\end{center}
\caption{SNR contours for the cross power spectrum in redshift space at the fiducial redshift 
$z= 2.5$ for a $400 \rm hrs$ observation at $405 \rm MHz$ with $10\%$ (left) and $100\%$ (right) 
foreground residuals remaining in the 21-cm signal.}
\label{fig:cross-residual}
\end{figure}

The right panel of the figure (\ref{fig:Hipps-nofg}) shows the SNR contours in the $(k_{||},
k_{\perp})$ plane for the Ly-$\alpha$ forest 21-cm cross-correlation power
spectrum. For the 21 cm signal, a $400 \, \rm hrs$ observation is
considered. We have considered a QSO number density of $\bar n = 30
\rm deg^{-2}$, and the Ly-$\alpha$ spectra are assumed to be measured
at a $2\sigma$ sensitivity level. We use $\beta_F$ to be $1.11$ and
overall normalization factor $C_F=-0.15$ consistent with recent
measurement \cite{slosar2011}.  The spectra is assumed to be smoothed to
the same level as the frequency channel width of the 21 cm
observations and the cross correlation is computed in the region of
overlap between the two fields. Although the overall SNR for the cross
correlation power spectrum is lower as compared to the 21-cm auto
correlation power spectrum, a $5\sigma$ detection is possible in the range
$0.1 \lesssim k_{\perp} \lesssim 0.4 \, {\rm Mpc^{-1}}$ and $0.1
\lesssim k_{\parallel} \lesssim 1 \, {\rm Mpc^{-1}}$. The SNR peaks ($>10$) at
$(k_{\perp}, k_{\parallel})\sim (0.2,0.3) \, {\rm Mpc^{-1}}$. We find
that in the variance budget $\sigma^2_{\hat{\mathcal{E}}}$
(eq. \ref{eq:sigma}) the second term always dominates over the first
term and therefore the variance is essentially determined by the
second term. In an ideal situation when the system noise for HI 21-cm
($N_T$) and Ly-$\alpha$ forest ($N_\F$) are zero or negligible, and $\bar{n}$ is large 
such that the term $P^{\rm {1D}}_{\F
  \F}(k_{\parallel}) P^{\rm {2D}}_{w} $ becomes very small, the two terms in the rhs. 
  of eq. \ref{eq:sigma} become comparable. 
This is the cosmic variance limited. However, in practice the second term is always significantly
larger than the first term except at very large scales. We further
notice that the second term, which dominates the variance budget in
almost all scales, is actually determined by the $P^{\rm {1D}}_{\F
  \F}(k_{\parallel}) P^{\rm {2D}}_{w} $ (which arises due to the
discreteness of QSO sight-lines) and the system noise in HI 21-cm survey
i.e, $N_T$. To be precise, we find that $P^{\rm {1D}}_{\F
  \F}(k_{\parallel}) P^{\rm {2D}}_{w} $ dominates  over
$P_{\F\F}({\bf k})$ for the scales $(k_{\perp}, k_{\parallel}) \gtrsim
(0.05,0.1) \, {\rm Mpc^{-1}}$. Similarly $N_T$ is higher than $P_{TT}$
for the scales $(k_{\perp}, k_{\parallel}) \gtrsim (0.2,0.3) \, {\rm
  Mpc^{-1}}$. The variance can be reduced either by increasing the QSO
number density or by increasing the observing time for HI 21-cm survey. The QSO
number density considered here is already on the higher side for the BOSS
like survey. Therefore, the only viable way to reduce the
variance is to consider more observation time for HI 21-cm survey.

It is important to note the role of foreground residuals in auto and
cross-correlation. Whereas foregrounds appear as an inseparable contaminant
to the auto-correlation signal, they appear only as a contribution to the
noise in the cross correlation. In the computation of SNR for
auto-correlation we have tacitly assumed that the foregrounds can be
distinguished from the true cosmological signal. This is practically
not possible and any foreground residual will plague the signal with
additional power which has no cosmological significance. The figure
\ref{fig:Hipps-nofg} is in fact hypothetical since it assumes that foregrounds
can be distinguished and completely separated from the signal. This issue is however not
present in the cross signal. Any foreground residual in the 21 cm
observation shall only degrade the noise and not affect the
cross correlated signal. This is a key advantage of using cross-correlation as a
cosmological probe as any detection here will ascertain the
cosmological origin of the signal. 

The figure \ref{fig:cross-residual} shows the cross-correlation in a
more realistic scenario wherein we consider the inclusion of
foreground residuals. Though the scale dependence of foreground residual may differ
significantly from that of the signal we have modeled foreground
residuals as merely $10\%$ and $100\%$ of the signal to get a rough
order of magnitude estimate about the degradation of SNR in the
presence of foreground residuals. We note that the $k-$dependence of the
contours here significantly depends on the nature and clustering
properties of the foregrounds and is also dependent on the foreground
subtraction technique.  We find that a peak SNR of $16$ and $12$
respectively can be achieved for these cases respectively.  We find
that the degradation of the peak SNR is not significant when the
foreground residual is $10\%$ of the signal. However, we note that
(see the right panel of Figure \ref{fig:cross-residual}) when
foreground residuals are as large as $100\%$ of the signal, there is
approximately $35\%$ reduction of the SNR as compared to the
no-foreground case.  The presence of a $100 \%$ foreground residuals
will clearly inhibit its detection using the 21-cm auto-correlation
power spectrum. However, we can see that a statistically significant
detection with high SNR is possible using the cross-correlation even if the foreground residual is 
$\sim 100\%$ of the HI 21-cm signal.

\subsection{Constraining parameters}

\begin{figure}[h]
\includegraphics[width=15cm, angle=0]{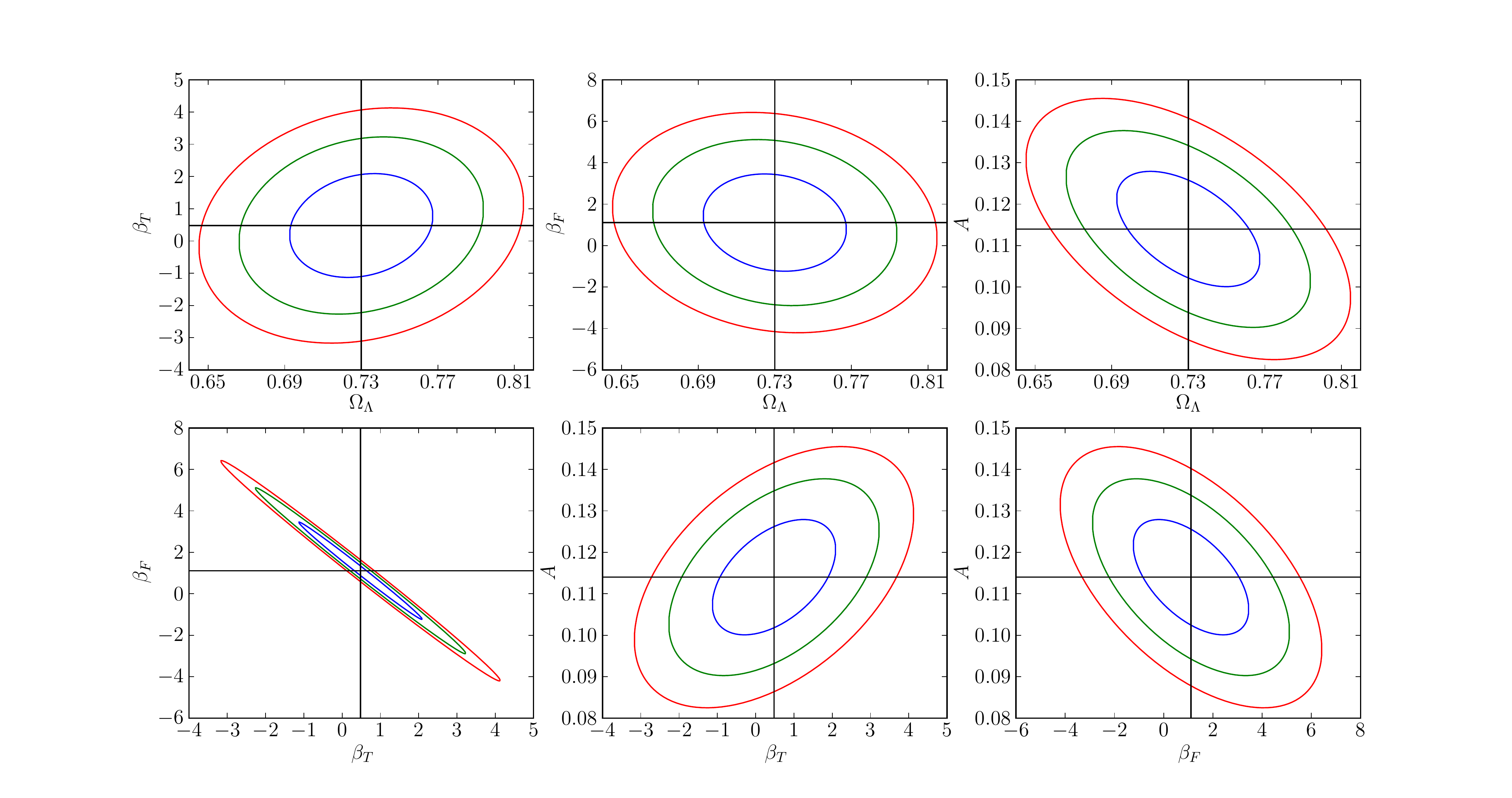}
\caption{Showing the $68.3 \%, 95.4 \%$ and $99.8\%$ confidence
  contours for the parameters $(A, \beta_{T}, \beta_{\F},
  \Omega_{\Lambda})$.}
\label{fig:ellipse}
\end{figure}
We shall now consider the possibility of constraining various model parameters
using the Fisher matrix analysis.
Figure (\ref{fig:ellipse}) shows the $68.3 \%, 95.4 \%$ and $99.8\%$
confidence contours obtained using the Fisher matrix analysis for the
parameters $(A, \beta_{T}, \beta_{\F}, \Omega_{\Lambda})$. The table \ref{tab:ska1-low} shows 
the $1-\sigma$ error for the above parameters. Owing to
the smallness of the contribution of the redshift space distortion to
the power spectrum, the parameters $\beta_{\F}$ and $\beta_{T}$ are
rather badly constrained and errors are very large. The parameters
$(\Omega_{\Lambda}, A)$ are constrained much better at $(3.5 \%, 8 \%)$
respectively. The  parameter $A$ is proportional to $b_T x_{\rm HI} C_{\F}$. Hence, the constraint on
$A$  is implicitly
related to the constraint on the mean neutral fraction. The
projections presented here are for a single field of view radio
observation.

\begin{figure}[h]
\includegraphics[width=7cm, angle=0]{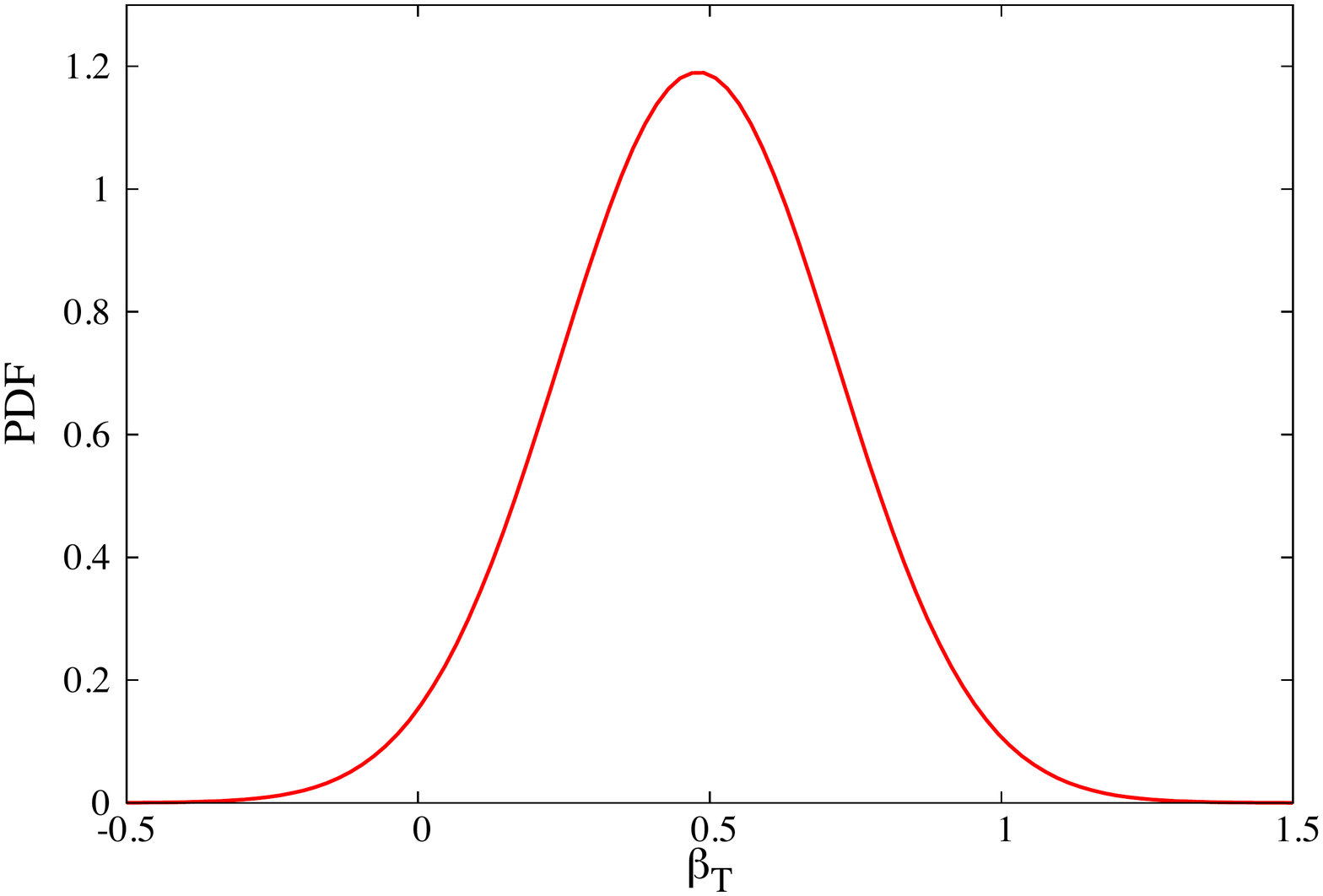}
\includegraphics[width=7cm, angle=0]{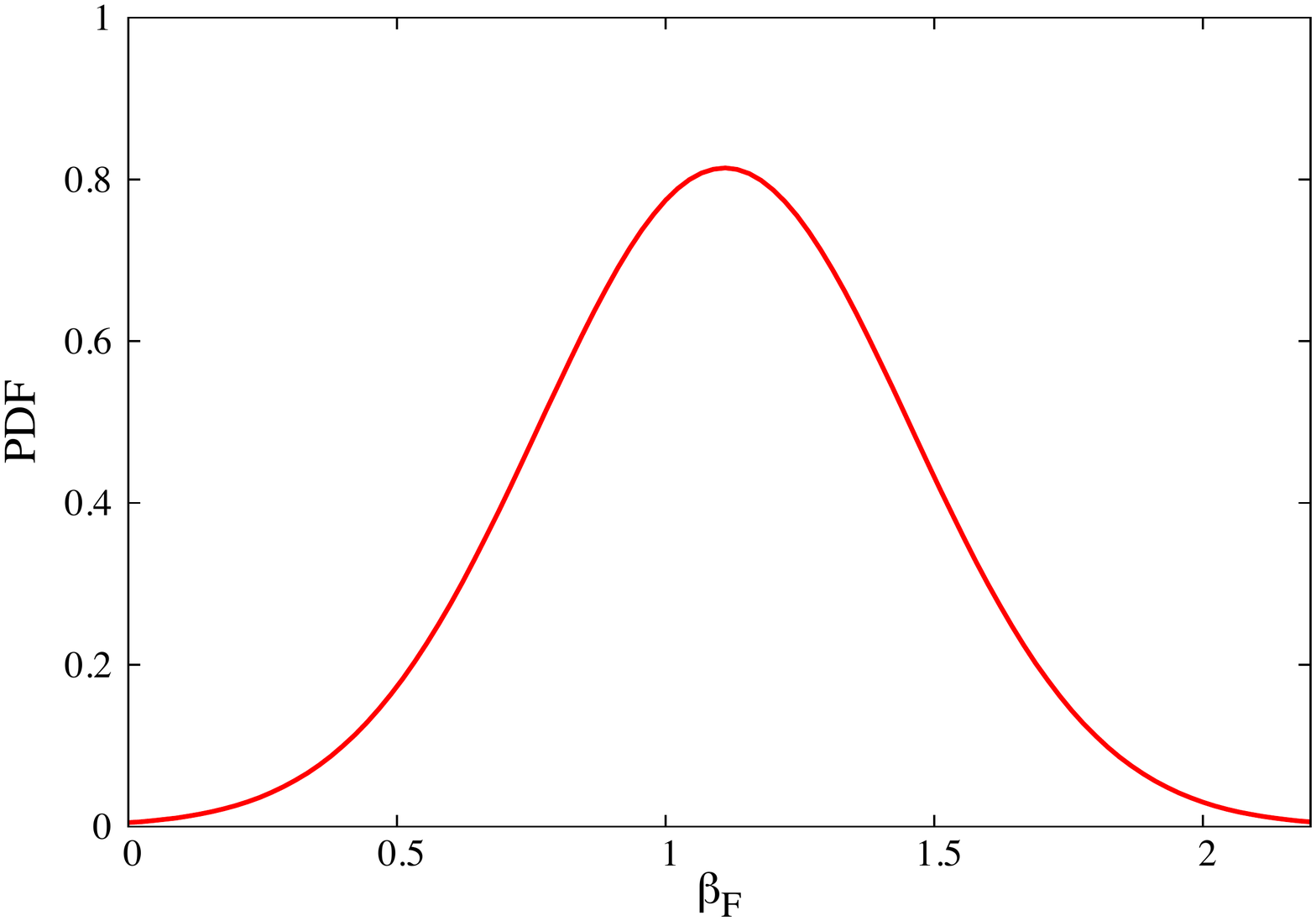}
\caption{Figure shows the marginalized one dimensional probability
distribution function (PDF) for $\beta_T$ and $\beta_F$ corresponding to $10$ pointings.}
\label{fig:pdf}
\end{figure}

The full coverage of a typical
Ly-$\alpha$ survey covers much larger volume and several beams of the
radio observation shall fit in this. The noise scales as $ \sigma/
\sqrt N$ where $N$ is the number of pointings. Such enhancement of
the volume shall allow the constraints to get much tighter. 
Figure \ref{fig:pdf} shows the marginalized one dimensional probability
distribution function (PDF) for $\beta_T$ and $\beta_F$ corresponding to $10$ pointings.
This gives an error on the parameters which are roughly of the same order of magnitude
as the fiducial values itself. The marginalized errors can however be reduced by considering more 
pointings.

\begin{table}[ht]
\caption{This shows $1-\sigma$ error on various parameters for a single field observation.} 
\centering 
\vspace{.2in}
\begin{tabular}{crrr} 
\hline
\hline 
 Parameters  & Fiducial Value & $1 \sigma$ Error &  $1 \sigma$ Error\\ 
&&(marginalized)&(conditional)\\
\hline
$\beta_{T}$ & 0.48 & 1.06 & 0.04 \\
\hline
$\beta_{\F}$ & 1.11 & 1.55 & 0.05 \\
\hline
$\Omega_{\Lambda}$ & 0.73 & 0.025 & 0.013\\
\hline
$A$ & 0.114 & 0.01 & 0.002\\
\hline
\end{tabular}
\label{tab:ska1-low}
\end{table}

Till now we have assumed that all the four 
parameters $(A, \beta_{T}, \beta_{\F}, \Omega_{\Lambda})$ are unknown. In the estimation 
of the parameters we have treated them as four  
free parameters. We find that the constraints on $\beta_T$ and $\beta_{\F}$ are
rather poor even if we consider $10$ independent pointings of radio observations.  We now consider error on each parameter assuming 
that the other three are known. This gives us the conditional  error on each parameter. The
conditional  $1-\sigma$ error in $\beta_T$ and $\beta_{\F}$ are $8.5
\%$ and $4.5\%$ respectively for single pointing radio observation.  For $10$ independent radio observations 
the conditional errors improve to $2.7 \%$, $1.4\%$, $0.4\%$ and $0.6\%$ for $\beta_T$,
$\beta_{\F}$, $\Omega_{\Lambda}$ and $A$ respectively. We note that these conditional errors 
give the best theoretical bounds on the parameters for the given observational specifications.
These  constraints obtained on redshift space distortion parameters $\beta$ from our cross-correlation analysis is
better as compared to the existing constraints \cite{andreu2,slosar2011} and competitive with 
other cosmological probes aiming towards the same
measurements. Further, higher density of QSOs and enhanced
SNR for the individual QSO spectra shall also ensure more stringent
constraints.

\section{Discussion \& Conclusions}

\begin{figure}[h]
\includegraphics[width=8cm, angle=0]{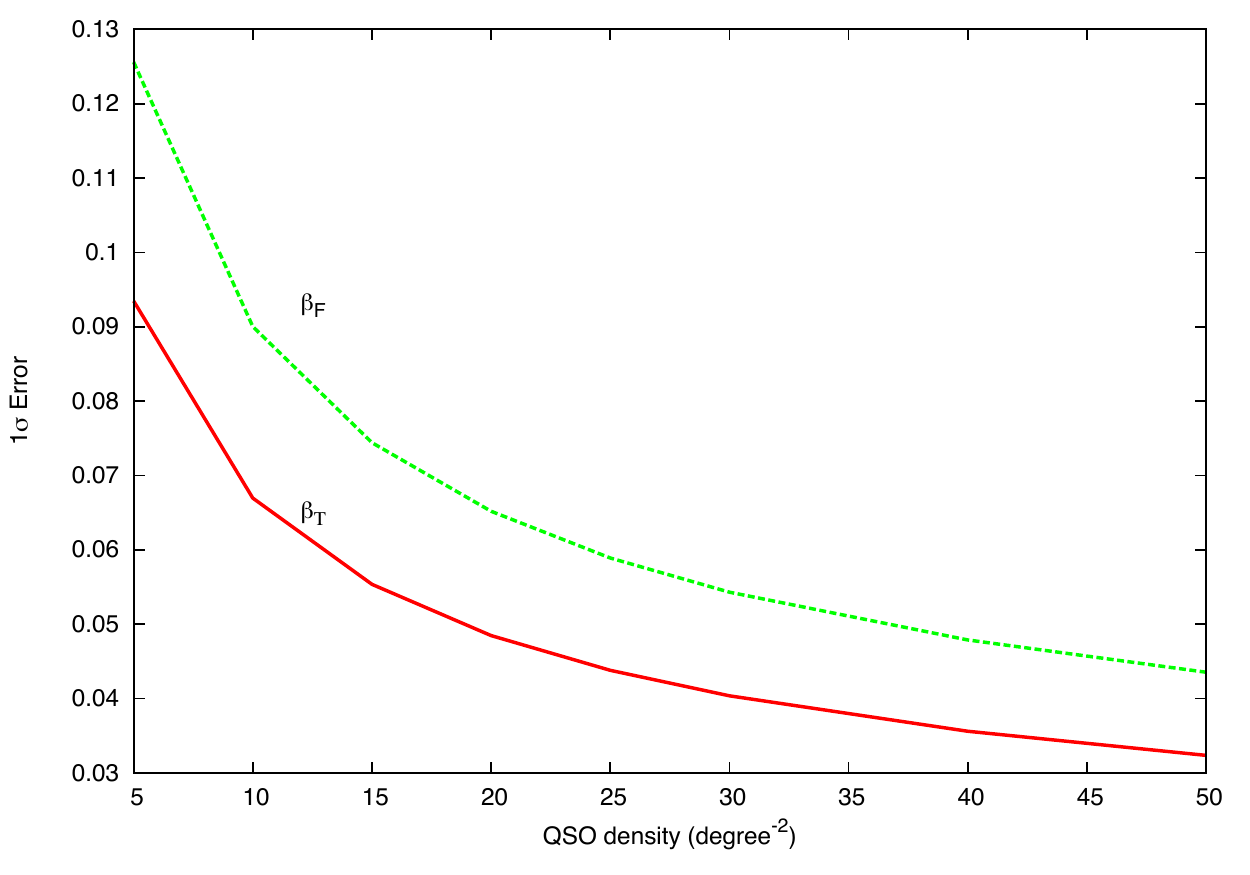}
\includegraphics[width=8cm, angle=0]{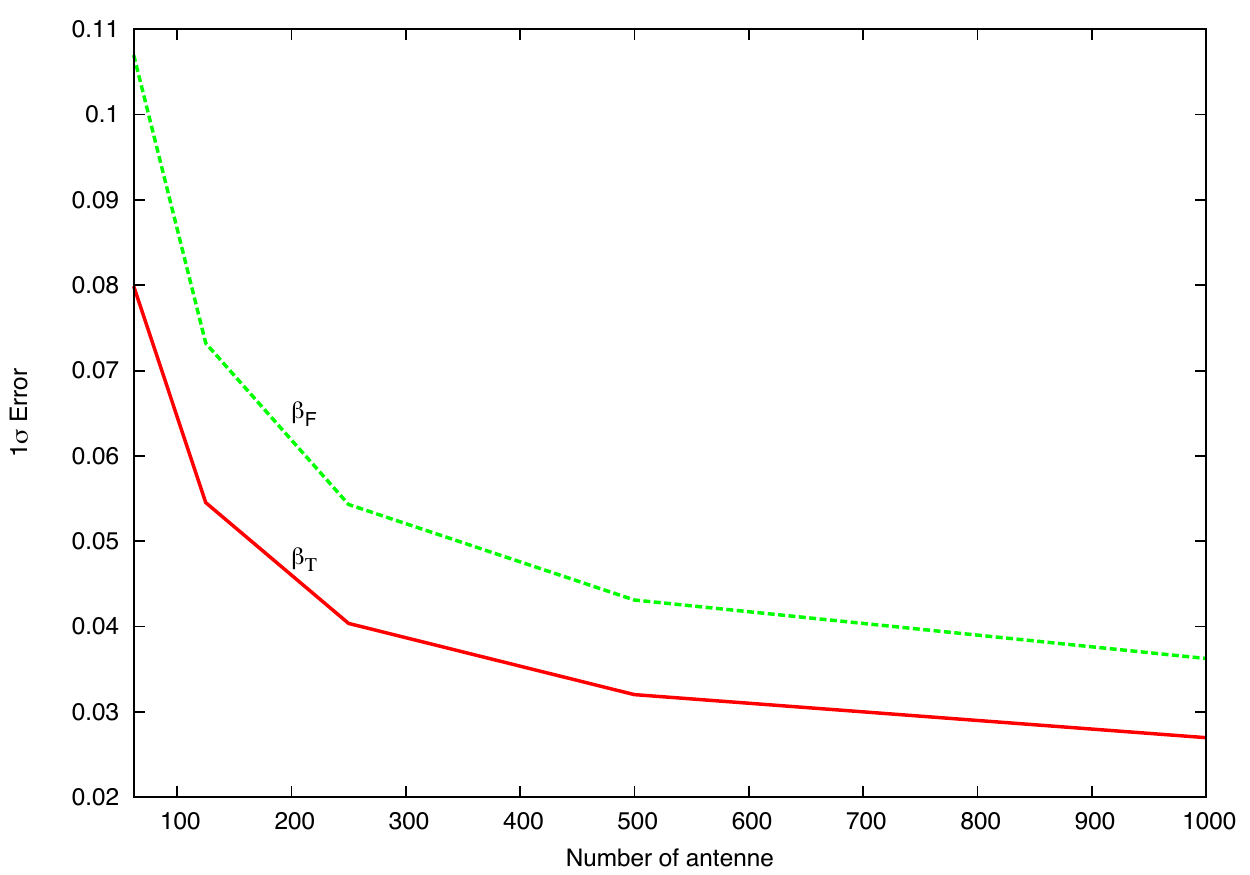}
\caption{ The left and right panels show 1$\sigma$ conditional errors on parameters $\beta_T$ and $\beta_F$ 
as a function of QSO number density ${\bar n}$ (left) and number of antennae (right) for total observing time $t_{obs}=400$ hrs 
and ${\bar n}=30 \, {\rm deg^{-2}}$  respectively for a single field  radio observation. We note that the total collecting area of the telescope increases linearly 
with increase in the number of antennae. All other parameters are kept fixed.}
\label{fig:nbar-nante}
\end{figure}

Our analysis has so far focused on estimates obtained for specific HI 21-cm intensity mapping 
(with SKA1-mid) and Ly-$\alpha$ forest (with BOSS) experiments. We shall now discuss how these estimates 
vary for various other possible experiments. 

We first consider the effect of QSO number density on the constraints on the redshift space distortion parameters. 
Fig \ref{fig:nbar-nante} (left) shows 1$\sigma$ conditional errors on redshift space distortion parameters $\beta_T$ and $\beta_F$ 
as a function of QSO number density ${\bar n}$ for a total observing time $t_{obs}=400$ hrs for a single field radio observation. 
All other observational parameters are held fixed. We find that there is a significant improvement
in the constraints with increase in the QSO number density. This improvement in the constraints 
is expected to saturate at very high
QSO number density. However, obtaining QSO number density ${\bar n} >50 \, {\rm deg^{-2}}$ with high SNR is unfeasible 
in near future and therefore we do not explore that possibility. The right panel of 
Fig. \ref{fig:nbar-nante} shows 1$\sigma$ conditional 
errors on parameters $\beta_T$ and $\beta_F$  as a function of number of antennae for
${\bar n}=30 \, {\rm deg^{-2}}$.  The total collecting area of the telescope increases linearly 
with increase in the number of antennae. The SKA telescope is in design phase and any degradation 
of baseline distribution through reduction of the proposed number of antennae will affect 
the constraints on $\beta_T$ and $\beta_F$ according to the Figure \ref{fig:nbar-nante}. We find a considerable improvement in the constraints
when the total number of antennae increases from $62$ to $500$, beyond which we hardly notice 
any improvement. Based on this, we argue that an increase in the number of antennae beyond $500$ for the SKA1-mid like
experiments is unlikely to improve constraints  on redshift space distortion parameters. 
We note that the SKA telescope is in design phase and any degradation 
of baseline distribution through reduction of the proposed number of antennae will affect 
the constraints on $\beta_T$ and $\beta_F$ according to the Figure \ref{fig:nbar-nante}.

\begin{table}[ht]
\caption{This shows $1-\sigma$ {\bf conditional} error on various parameters for different
 baseline distributions.} 
\centering 
\vspace{.2in}
\begin{tabular}{crrrr} 
\hline
\hline 
 Parameters  &  SKA1-mid&  n=1& n=2 & n=3\\ 
&&&\\
\hline
$\beta_{T}$ & 0.04 & 0.037 & 0.032 &  0.032 \\
\hline
$\beta_{\F}$ & 0.05 & 0.049 & 0.044 & 0.044\\
\hline
$\Omega_{\Lambda}$ & 0.013 & 0.012 & 0.011& 0.011\\
\hline
$A$ & 0.002 & 0.002 & 0.0018  & 0.0018\\
\hline
\end{tabular}
\label{tab:baseline}
\end{table}

\begin{figure}[h]
\begin{center}
\includegraphics[width=6cm, angle=270]{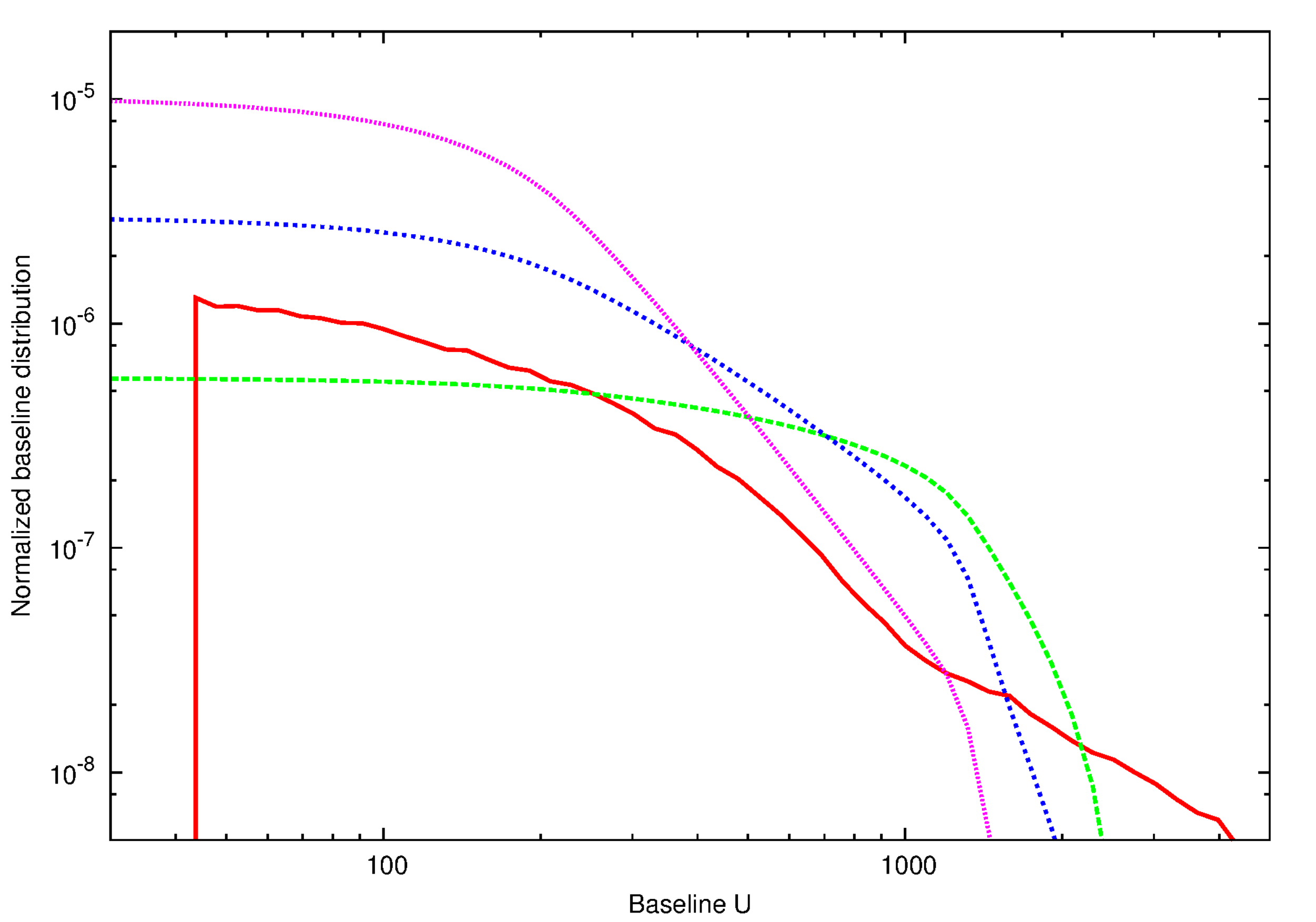}
\caption{The normalized baseline distribution function $f_{\rm 2D}(U,\nu)$. The solid curve shows
the normalized baseline distribution for the SKA1-mid. The other three curves correspond to power law antenna
distribution with power law indices $n=3 \, , 2\, , 1$ (top to bottom). }
\label{fig:baseline-dis}
\end{center}
\end{figure}

An important feature of radio interferometer  design is the array configuration which determines the 
baseline distribution $f_{2D}(U, \nu)$. In the previous section we considered the baseline 
distribution function for the SKA1-mid. We now consider three other baseline distribution 
corresponding to radially symmetric antenna distribution of the form $\sim r^{-n}$ with 
$n=1 \, , 2\, , 3$. We assume that the array has an uniform antenna distribution up to a radius 
of $80$m. We also assume that all the antennae are confined within a radius of $1$ km. 
Fig. \ref{fig:baseline-dis} shows the normalized baseline distribution function for these cases.

We wish to study the role of the compactness of the radio array on parameter estimation. The 
1$\sigma$ conditional errors for these baseline distribution functions are summarized in 
Table \ref{tab:baseline}. We find that there is a general improvement of our results as compared to 
the SKA1-mid when all the antennae are confined within a radius of $1$ km. We note that for 
the SKA1-mid, about $ 50 \%$ of the antennae are outside the $1$ km core where as all antennae
are confined within a radius of $1$ km for the above three array designs. We find that there 
is a slight improvement in the constraints when the power law index is varied from 
$n=1$ to $2$. But, hardly any improvement is noticed when the index is changed from $n=2$ 
to $n=3$. 

It is clear that distributing antennae at large distances in such experiments (where resolution is 
not a prime concern ) is not advantageous since at large baselines the signal is sub-dominant. 
On the contrary it is useful to consider arrays in which most antennae packed within a small
radius. However, we note that an arbitrary compactification (from $n=2$ to $3$)
of the array causes us to lose $k-$modes which contribute to the signal and there is no further improvement in the SNR.

It is important to choose the optimal observational strategy for the 21-cm signal.
One may consider a deep observation (long time of observation) in a single field of view 
as opposed to the possibility of dividing the  same total  observation time  over many pointings.
We know that on large scales the dominant contribution to noise comes from the cosmic variance. This 
can only be mitigated by increasing the survey volume. Deep observations in a small patch 
of the sky is not beneficial for reducing the cosmic variance. Observing multiple fields
ensures that both the system noise contribution and the cosmic variance for the cross-correlation 
is reduced. The noise in this case is reduced on all scales. However, if one considers an observation 
for the same total time but in a single field of view, the system noise contribution 
indeed reaches the same minimum value but the noise which is dominated by the cosmic variance on large scales does not get reduced.
It is thus strategically better for cross-correlation measurements to consider observations in multiple fields.

The issue of foreground subtraction, though less severe for the
cross-correlation is still a major concern for the 21-cm
signal. Astrophysical foregrounds from galactic and extra galactic
sources plague the signal \cite{ghosh2011} and significant amount of
foreground subtraction is required before the cross-correlation is
performed.  The foregrounds appear as noise in the cross-correlation
and may be tackled by considering larger number of Fourier modes
(larger volumes).  Similarly, for the Ly-$\alpha$ forest observations, continuum
subtraction and avoiding metal line contamination, though less
problematic, has to be performed with extreme precision. Moreover, man made radio frequency 
interferences (RFIs), calibration errors
and other systematics pose a serious threat to the detection of the HI 21-cm signal. 
A detailed analysis of these issues is outside the scope of the present work. We intent to 
address these observational aspects in a future work.

Finally, this work emphasizes the important role of using cross correlation to
bypass the fundamental problem posed by 21-cm foregrounds towards its
detection through auto-correlation. We have shown that the 3D cross-correlation power spectrum
from the post reionization epoch as a direct probe of cosmological
structure formation can be detected to a high level of statistical
sensitivity with telescopes such as the SKA. We have also focused on
the possibility of constraining the bias (redshift space distortion)
parameters for the 21-cm signal and the Ly-$\alpha$ forest. This
investigation is crucial towards understanding the nature of HI bias
in two distinct astrophysical systems under consideration namely
Ly-$\alpha$ forest (diffuse low density HI in the IGM) and 21-cm
signal (clumped HI in DLAs). These biases are studied extensively in
numerical simulations. It is important for observational constraints to
be compared with the simulation results to support their validity. 
Modeling of the IGM is incomplete without  the precise knowledge about HI bias.
In the absence of high
quality 21-cm data it is important to make predictions based on future
experiments. We find that strong constraints may be obtained on
$(\beta_T, \beta_{\F})$ from advanced BOSS and SKA like experiments.

We conclude by noting that the cross-correlation of the Ly-$\alpha$
forest and the HI 21 cm signal as an independent probe of astrophysics  and cosmology may
allow us to put strong constraints on redshift space distortion parameters
towards important understanding and modeling of the post reionization HI
distribution.

\section*{Acknowledgements}
The authors would like to thank Anjan Sarkar for useful discussion on Fisher matrix analysis.
The authors would also like to thank Mario G. Santos for providing them with the SKA1-mid array 
baseline distribution. TGS would like to acknowledge the Department of Science and Technology
(DST), Government of India for providing financial support through the
projects SR/FTP/PS-172/2012. KKD would like to thank DST for support
through the project SR/FTP/PS-119/2012 and the University Grant
Commission (UGC), India for support through UGC-faculty recharge scheme
(UGC-FRP) vide ref. no. F.4-5(137-FRP)/2014(BSR).

\bibliographystyle{JHEP}
\bibliography{ms1_m.bbl}

\end{document}